\documentclass[prl, twocolumn,showpacs,preprintnumbers,amsmath,amssymb,a4paper]{revtex4}
\usepackage{graphicx}
\newcommand{\imag}{\mathrm{i}}

\begin{document}

\title{Nonlocality-induced front interaction enhancement}

\author{L.~Gelens$^{1,2}$, D.~Gomila$^{2}$, G.~Van~der~Sande$^1$,  
M.~A.~Mat\'{\i}as$^2$, and P.~Colet$^2$}
\affiliation{
$^1$Department of Applied Physics and Photonics, Vrije Universiteit Brussel,
Pleinlaan 2, 1050 Brussels, Belgium;\\
$^2$IFISC, Instituto de F\'{\i}sica Interdisciplinar y Sistemas 
Complejos (CSIC-UIB), Campus Universitat Illes Balears, E-07122 Palma de
Mallorca,  Spain
}



\begin{abstract} 
We demonstrate that nonlocal coupling strongly influences the dynamics of
fronts connecting two equivalent states. In two prototype models we observe a
large amplification in the interaction strength between two opposite fronts
increasing front velocities several orders of magnitude. By analyzing the spatial dynamics we prove that way beyond
quantitative effects, nonlocal terms can also change the overall qualitative
picture by inducing oscillations in the front profile. This leads to a
mechanism for the formation of localized structures not present for local
interactions. Finally, nonlocal coupling can induce a steep broadening of
localized structures, eventually annihilating them.
\end{abstract}

\maketitle

Most studies on the emergence of complex behavior in spatially extended systems 
consider that spatial coupling is either local or alternatively global (all to
all  coupling) \cite{Kuramoto_Springer_1984,Cross_RevModPh_1993}. More recently,
systems  with nonlocal (or intermediate- to long-range) coupling have received
increasing attention, as nonlocal interactions are known to be relevant in
diverse fields, ranging from Josephson junction arrays \cite{Phillips_PRB_1993}
and chemical reactions \cite{Kuramoto_PRL_1998, Shima_PRE_2004}, to several
problems in Biology and Ecology \cite{Murray_Springer_2002}, such as the neural
networks underlying mollusk patterns \cite{Ermentrout_Shells,coombes05}, ocular
dominance stripes and hallucination patterns \cite{Ermentrout_RPP},  and
population dynamics  \cite{2004PhRvE_Hernandez}. A nonlocal interaction may
emerge from a physical/chemical mechanism that couples points far apart in
space, e.g., a long-range interaction \cite{PedriSantos}, or from the adiabatic
elimination of a slow variable \cite{Kuramoto_PTP_1995,Kuramoto_PRE_2003}. Novel
phenomena emerging genuinely from nonlocality, such as power-law correlations
\cite{Kuramoto_PTP_1995,Kuramoto_PRL_1996},  multiaffine turbulence
\cite{Kuramoto_PRL_1998}, and chimera states  \cite{Abrams_PRL_2004,
Sethia_PRL_2008} have been reported. Moreover, recent works have reported the
effects of nonlocality on the dynamics of fronts, patterns and localized 
structures (LS), for instance the tilting of snaking bifurcation lines 
\cite{Firth_PRL_2007} and changes in the size of LS \cite{Gelens_PRA}, the
effects of two-point nonlocality on convective instabilities 
\cite{Zambrini2005}, the nonlocal stabilization of vortex beams
\cite{Skupin-2007},  or changes in the interaction between solitons \cite{Neshev},
and in the velocity of propagating fronts \cite{Maruvka2007}. 

The main goal of this Letter is to show the crucial relevance of 
nonlocality on the interaction of fronts connecting two equivalent states in 
one dimensional systems, as well as on the formation of LS arising from the 
interaction of two such fronts \cite{Pesch_PRL_2007}. Interaction between two 
monotonic fronts is always attractive, so any domain of one state embedded in 
the other shrinks and disappears. However, fronts with oscillatory tails can 
lock at specific distances leading to stable LS. Here we show that oscillatory 
tails, and therefore stable LS can appear as an effect of repulsive 
nonlocal interactions. Repulsive (inhibitory) interactions are common, for
instance, in neural field theories \cite{coombes05, Ermentrout_RPP} and genetic networks
\cite{Alonbook}. Our result is generic and can be qualitatively understood from
the interplay between nonlocality, which couples both sides of the front, and 
repulsiveness which induces a small depression at the lower side and a small 
hill at the upper part. Altogether this leads to oscillatory tails as qualitatively obtained from the spatial dynamics.

A prototypical model of a spatially extended system with two equivalent steady
states is the real Ginzburg-Landau equation (GLE) \cite{Cross_RevModPh_1993, 
Aranson_RevModPh_2002}. We consider the 1D {\it nonlocal} real GLE
\begin{equation}
	\frac{\partial E(x)}{\partial{t}}  = (\mu - s) E(x) +  
	\frac{\partial^2 E(x)}{\partial{x^2}} - E^3(x)  + 
	s F(x) ,
\label{Eq::RGL} 
\end{equation}
being $E(x)$ a real field and $\mu$ the gain parameter. The diffusion and the nonlinear term have been scaled to one without loss of generality. Eq. (\ref{Eq::RGL}) has
both a local (diffusive) and a (linear) nonlocal spatial coupling 
\begin{equation}
F(x) =   \int_{-\infty}^{\infty} \theta_{\sigma}(x-x')E(x')dx' ,
\label{Eq::Fnloc}
\end{equation}
where $\theta_{\sigma}$ is the spatial nonlocal interaction function (or kernel)
and $\sigma$ indicates the spatial extension of the coupling. For the sake of simplicity, we consider here a
Gaussian kernel, $\theta_{\sigma}(x-x') = (1/ \sqrt{2 \pi} \sigma) e^{-(x-x')^2
/ (2 \sigma^2)}$, although the results presented in this work do not depend
qualitatively on its precise shape, provided it is positively defined. Gaussian
kernels appear in contexts such as mollusk pigmentation patterns
\cite{Ermentrout_PNAS} and Neuroscience \cite{Ermentrout_RPP,Hellwig}. The nonlocal function $F(x)$ includes also a local contribution. This contribution is compensated for by the term $-s E(x)$, such that in the limit $\sigma \rightarrow 0$ one recovers the same results as for the local GLE.

\begin{figure}
\begin{center}
\includegraphics[width=7.5cm]{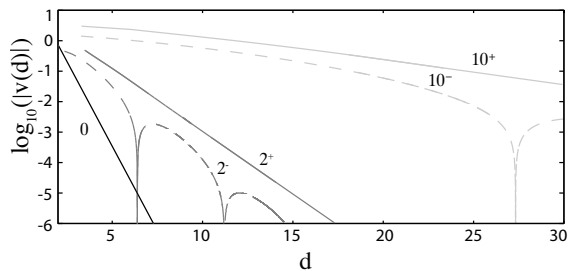}
\caption{Front velocity as a function of 
$d$. $\mu = 3, \vert s \vert=1$, and $\sigma$ is taken to be $0$ , $2$ and $10$, depicted in the
figure as $\sigma^{\text{sgn($s$)}}$.}
\label{Fig::log10veloc_width_sigma_2_10}
\end{center}
\end{figure}

The profile of GLE fronts is known analytically \cite{Coullet87}, and the interaction with an opposite front located at a distance
$d$ much larger than the front width can be calculated perturbatively, yielding
the following result for the relative velocity $v(d)$ \cite{Coullet87},
\begin{equation}
v(d)  =  \dot{d} = c\, e^{- \gamma d} \text{ ,} \label{Eq::expLaw} 
\end{equation}
with $c =  -24 \sqrt{2 \mu}$ and $\gamma = \sqrt{2 \mu}$. 
The nonlocal effects in front dynamics can be quantified by 
looking at the deviations from the interaction given by Eq.\ (\ref{Eq::expLaw}),
stemming solely from a local interaction coupling. Therefore we have studied the
front velocity $v(d)$ for different kernel widths $\sigma$ keeping all other
parameters fixed and taking the system  size much larger than $\sigma$.  The
results are plotted in Fig. \ref{Fig::log10veloc_width_sigma_2_10}. The {\it
local\/} case, Eq.\ (\ref{Eq::expLaw}), is the straight line labeled with $0$
(as $\sigma=0$). In the nonlocal case ($\sigma \ne 0$), a
large  qualitative difference in behavior is observed depending on the sign of
$s$.

\begin{figure}
\begin{center}
\includegraphics[width=7.5cm]{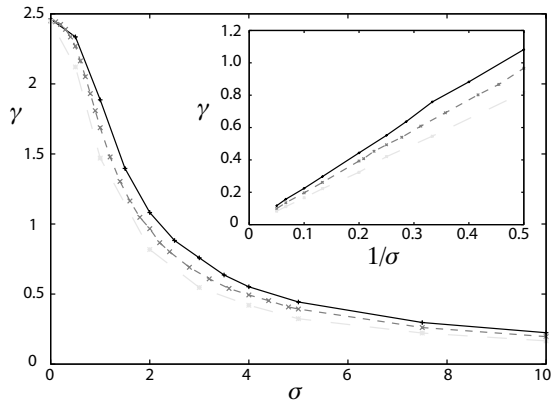}
\caption{$\gamma$ as a function of $\sigma$. 
Inset: $\gamma$ as a function of $1/\sigma$. The black solid, dark gray
dashed and light gray dashed curves have $\mu = 3$, and $s = 0.5$, $1$, and $2$,
respectively.}
\label{Fig::Sigma_Gamma_lin_sPos}
\end{center}
\end{figure}

\begin{figure}
\begin{center}
\includegraphics[width=7.5cm]{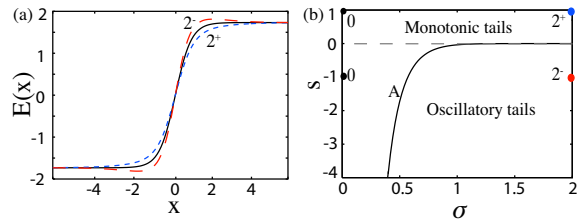}
\caption{(Color online) Nonlocal GLE, Eq.~(\ref{Eq::RGL}) with $\mu = 3$: 
(a) Front profiles for $s=0$ (black line) and  $s=\pm 1$ with $\sigma =
2$, labeled in the figure as $\sigma^{\text{sgn($s$)}}$. (b) The
curve $A$ is the boundary in the $(\sigma, s)$ space between the presence
of oscillatory or monotonic tails of a front.}
\label{Fig::SpatialEigBoundary_Sigma_S_realGLE}
\end{center}
\end{figure}

As shown in Fig.\ \ref{Fig::log10veloc_width_sigma_2_10}, for an attractive (activatory) interaction ($s>0$), the logarithm of the velocity decreases linearly with the distance, such that the exponential dependence of the velocity with the distance given by Eq.\
(\ref{Eq::expLaw}) still holds with an effective $\gamma$ whose value is strongly reduced by the nonlocality. As a consequence the range of spatial interaction, $1/\gamma$, increases with the kernel width $\sigma$. Even for moderate values of $\sigma$, fronts move several orders of magnitude faster than for the local GLE. Fig.\ \ref{Fig::Sigma_Gamma_lin_sPos} shows the change of the value of the effective exponent $\gamma$ as a function of $\sigma$. Moreover, the inset shows that $\gamma$, to a very good extent, follows a linear dependence with the inverse of the kernel width $\sigma$, provided that $\sigma$ is at least as large as the front width. Therefore, we can conclude that rescaling $\gamma$ to $\gamma/\sigma$ the effective interaction between two fronts follows a universal exponential law (except for a small
dependence on the strength of interaction $s$). It is interesting to notice,
however, that the width of the front, defined by the  half width at half the
maximum (HWHM) does not show this scaling with $1/\sigma$ cf. 
Fig.~\ref{Fig::SpatialEigBoundary_Sigma_S_realGLE}(a).  This can be understood
by noticing that while the general shape of the front is  mainly dominated by
the local, diffusive, coupling, in turn the nonlocal coupling modifies
substantially the exponential tails, which are responsible of the long-range
interaction that influences the front velocity.

For $s < 0$, i.e., for a repulsive (inhibitory) interaction, the exponential
law, Eq.\ (\ref{Eq::expLaw}), no longer holds. Nevertheless, the magnitude  of
the envelope of the front velocity is still dominated qualitatively by
(\ref{Eq::expLaw}), as shown in Fig.\ \ref{Fig::log10veloc_width_sigma_2_10}. In
this case, the velocity becomes zero at regular intervals of the distance $d$
between two fronts. At these positions the fronts are locked leading to the formation of LS. These
LS, not present in the GLE, come into existence by the creation of oscillatory
tails when a nonlocal term with $s < 0$ is added. 

\begin{figure}
\begin{center}
\includegraphics[width=7.5cm]{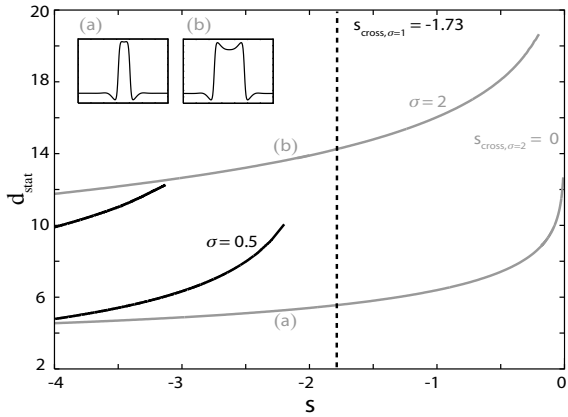}
\caption{\label{Fig::1D_bifdiag_RGLE} Stationary widths of stable LS for the nonlocal GLE ($\mu = 3$). Black (gray) 
line corresponds to $\sigma=0.5$ ($\sigma=2$). Insets show examples of LS profiles
for $\sigma=2$. $s_{cross,\sigma}$, given by line A in 
Fig.~\ref{Fig::SpatialEigBoundary_Sigma_S_realGLE}, indicates the point where the spatial eigenvalues go from being complex conjugate to purely real.}
\end{center}
\end{figure}

A quantitative characterization is obtained from the
{\it spatial  dynamics} of Eq.\
(\ref{Eq::RGL}). Considering a spatial perturbation to the homogeneous solution
of the form $E = \sqrt{\mu} + \epsilon \exp(\lambda x)$, one finds
\begin{equation}
-2 \mu - s + \lambda^2 + s e^{\lambda^2 \sigma^2/2} = 0. 
\label{Eq::spatialDyn_realGLE}  
\end{equation}
Eigenvalues come in pairs $\pm \lambda$. The shape of the front is determined 
by the eigenvalue $\lambda_1$ with real part closest to zero, as in the spatial dynamics 
all the other directions are damped faster when approaching the fixed point
(homogeneous solution).  
By determining where $\lambda_1$ goes from being purely real to complex, one 
can find the boundary $A$ separating fronts having monotonic and oscillatory
tails. As shown in Fig.~\ref{Fig::SpatialEigBoundary_Sigma_S_realGLE}(b), for kernel widths $\sigma$ at least as large as the front width, the front profile always shows oscillatory tails for $s<0$ and as a consequence LS can arise as displayed in Fig.
\ref{Fig::1D_bifdiag_RGLE}. For smaller kernel widths an increasingly large
nonlocal strength is needed for LS to be formed (see line A in Fig.
\ref{Fig::SpatialEigBoundary_Sigma_S_realGLE}). To describe the interaction of
fronts with oscillatory tails Eq.\ (\ref{Eq::expLaw}) must be modified. An
appropriate ansatz is the following:
\begin{equation} 
v  =  \dot{d} = c \cos[\zeta(s, \sigma) d] e^{- \gamma(s, \sigma) d} \text{ ,}
\label{Eq::expLaw_PGLE_1D}  
\end{equation} 
where $\zeta(s, \sigma)$ is determined by the complex part of the most
underdamped spatial eigenvalue of Eq.\ (\ref{Eq::spatialDyn_realGLE}), such that
$\zeta = 0$ in the region of monotonic tails. Eq.~(\ref{Eq::expLaw_PGLE_1D})
adequately describes the dependence of the velocity with the distance indicated in 
Fig.~\ref{Fig::log10veloc_width_sigma_2_10}. 

\begin{figure}
\begin{center}
\includegraphics[width=7.5cm]{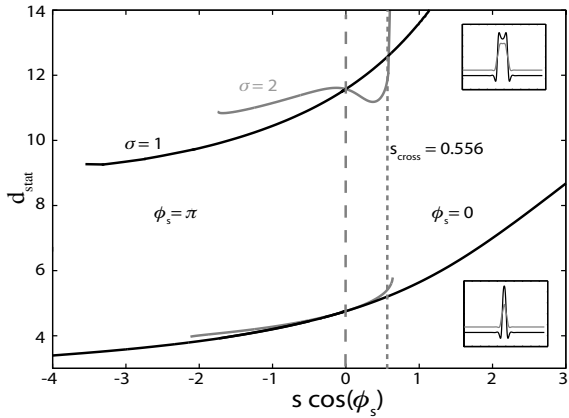}
\caption{\label{Fig::1D_snaking_sigma_1en2} Stationary widths of stable
LS for $\alpha = 1$, $\beta=0$, $\mu = 0$, $\nu=2$ and $p=2.7$.
$\sigma= 1$ (black
line) and $2$ (gray line). Insets show the transverse profile of the LS.}
\end{center}
\end{figure}

\begin{figure}
\begin{center}
\includegraphics[width=7.5cm]{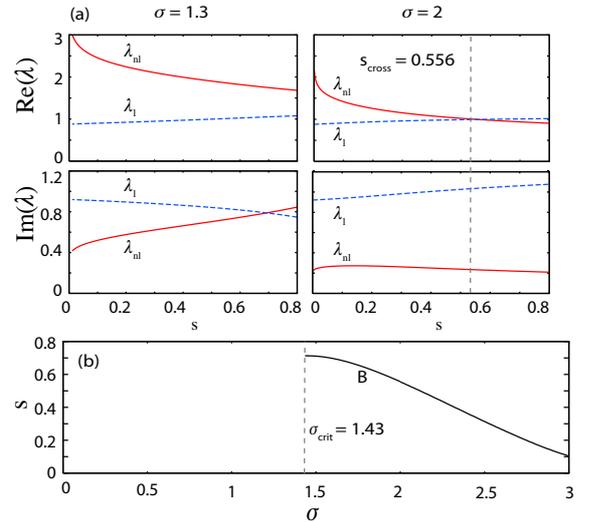}
\caption{\label{Fig::SpatialEigBoundary_sigma_s_phi_0_PCGLE} (Color online)  (a)
Dependence of the absolute value of the real and imaginary parts of the two
dominant complex quartets of spatial eigenvalues in the nonlocal PCGLE for $\sigma = 1.3$ and  $\sigma = 2$ as a function of $s$.
Other parameters as in Fig.~\ref{Fig::1D_snaking_sigma_1en2}. (b) 
The curve $B$ depicts the locus in the $(\sigma,  s)$ plane of the points where
the absolute values of the real parts of $\lambda_l$ and $\lambda_{nl}$ cross.}
\end{center}
\end{figure}

In summary, for the real GLE one finds an enhancement of the
interaction due to nonlocality, that leads to the appearance of oscillatory
tails and LS for a repulsive interaction and increase of the front velocity if
the interaction is attractive. 

We check the generality of these findings
for the Parametrically forced Complex Ginzburg Landau Equation
(PCGLE) whose fronts show oscillatory tails already in the local case
and form LS \cite{Gomila01, Yochelis2006}. 
The PCGLE is the generic amplitude equation for oscillatory systems 
parametrically forced at twice the natural frequency \cite{Coullet90}.
The nonlocal version of the PCGLE can be written as: 
\begin{eqnarray}
\frac{\partial E(x)}{\partial t}  = (1 + \imag \alpha) 
\frac{\partial^2 E(x)}{\partial{x^2}} 
+ [ (\mu +\imag \nu) - s e^{\imag \phi_s} ]E(x) \nonumber \\ 
- (1+ \imag\beta)\vert{E(x)}\vert^2\,E(x) + p E^{*}(x) 
+ s e^{\imag \phi_s} F(x),
\label{Eq:PCGLE}
\end{eqnarray} 
where $\mu$ measures the distance from the oscillatory instability threshold,
$\nu$ is the detuning between the driving and the natural frequencies, $p >
0$ is the forcing amplitude, $\alpha$ and $\beta$ represent the linear and
nonlinear dispersion. The term $s e^{\imag \phi_s}
F(x)$ describes the nonlocal response of the material taking
the same form as in the study of the GLE (\ref{Eq::Fnloc}). Here we consider $s>0$ with $\phi_s = 0$ and $\phi_s = \pi$ in correspondence with attractive and repulsive interactions discussed previously. Again, the linear
contribution has been compensated in the term $-s e^{\imag \phi_s}E(x)$.

Fig.\
\ref{Fig::1D_snaking_sigma_1en2} shows the width of the stable LS locked
at the first and second oscillations as a function of the nonlocal strength $s$.
The insets show the LS spatial profile. Whereas in the real GLE, LS existed only for $s<0$ (see
Fig.\ \ref{Fig::1D_bifdiag_RGLE}), in the PCGLE, LS also exist for a finite
range of positive values of $s \cos(\phi_s)$ $(\phi_s=0)$. This is shown
in Fig.\ \ref{Fig::1D_snaking_sigma_1en2} for $\sigma = 2$ where the bifurcation
branch abruptly ends around $s_{cross} = 0.556$. This can be explained again by studying
the spatial eigenvalues of the system.
In this case there are two pairs of complex conjugate eigenvalues with small
real part $\lambda_l$ and $\lambda_{nl}$ which play a relevant role.
Fig.~\ref{Fig::SpatialEigBoundary_sigma_s_phi_0_PCGLE}(a) shows the dependence 
of the real and imaginary parts of these two eigenvalues for
$\sigma = 1.3$ and $2$ as a function of $s$. 
For small $s$ the spatial dynamics is dominated by $\lambda_{l}$, an eigenvalue
already present in the local case. Increasing $s$ the real part of this
eigenvalue becomes slightly larger while the real part of $\lambda_{nl}$ 
clearly decreases. 
For $\sigma = 2$, the real parts of the two eigenvalues cross at $s_{cross}=0.556$. 
Beyond that value the spatial dynamics is governed by $\lambda_{nl}$ 
whose
spatial length scale [$\text{Im}(\lambda_{nl})$] is an order of magnitude larger 
than the one corresponding to $\lambda_{l}$. As a result, the branch of LS experiences dramatic sharpening in Fig.\ \ref{Fig::1D_snaking_sigma_1en2}, as the LS broaden with an order of
magnitude. 
Fig. \ref{Fig::SpatialEigBoundary_sigma_s_phi_0_PCGLE}(b) shows the locus
of the crossing point in the $(\sigma,  s)$ plane for $\phi_s = 0$. For large
values of $\sigma$,  the crossing point moves towards $s = 0$, while for $\sigma<1.43$ 
the crossing never takes place and the spatial dynamics is always dominated by $\lambda_l$.

For $\phi_s = \pi$, nonlocality only modifies slightly the
preexisting oscillatory tails of the PCGLE. The branches of LS end for negative
values of $s \cos(\phi_s)$ close to the modulational instability point of
the background states (that for $\sigma=2$ is at $s \cos(\phi_s) \sim -3$).
This MI point, that was originated by the nonlocality in the GLE, is already
present in the local form of the PCGLE but  is enhanced by the nonlocal
interactions.

In conclusion, we have demonstrated the large impact of a linear nonlocal term
on the interaction of fronts connecting two equivalent states. The most striking
result is the possibility of obtaining a novel class of self-organized stable
localized structures in systems exhibiting fronts with no tails, like the real
GLE. This is achieved through a nonlocal repulsive interaction that modifies the
profile of the fronts introducing or damping out oscillations. In addition, we
have observed an order of magnitude increase of the front velocity due to an
enhancement of the interaction between two fronts. Finally we have shown  that
nonlocal interactions can also smooth out the front oscillations greatly
increasing their  wavelength such that the LS become much wider and eventually
disappear. 

The characterization of the effects of nonlocal coupling on the front properties
and dynamics can allow the identification, from both theoretical and
experimental data, of different sources of nonlocality. For instance, domain
walls have long been studied in photorefractive media \cite{photorefractive}, 
which have a large nonlocal response, but its effects on the front dynamics 
have never been identified. Nonlocal interactions
are also common in Biology, Chemistry, and other fields, and they can have a
constructive role by enhancing the propagation of information between distant 
parts of the system.

This work was supported by the Belgian Science Policy Office and by
the Spanish MICINN and FEDER under grants No.\ IAP-VI10 and FIS2007-60327.
GVdS and LG acknowledge support by the
Research Foundation-Flanders (FWO). We thank Prof. E. Knobloch for interesting
discussions.

\end{document}